\documentclass[aps,prl,twocolumn,showpacs,preprintnumbers,amsmath,amssymb,floatfix,nofootinbib]{revtex4}
\usepackage{graphicx}
\usepackage{dcolumn}
\usepackage{bm}

\newcommand{\dd}{\mathrm{d}}

\newcommand{\w}{\wedge}
\newcommand{\bbm}{\left(\begin{matrix}}
\newcommand{\ebm}{\end{matrix}\right)}
\newcommand{\beq}{\begin{eqnarray}}
\newcommand{\eeq}{\end{eqnarray}}
\makeatother

\newcommand{\sfrac}[2]{{\textstyle\frac{#1}{#2}}}

\newcommand{\be}{\begin{equation}}
\newcommand{\ee}{\end{equation}}
\newcommand{\beqa}{\begin{eqnarray}}
\newcommand{\eeqa}{\end{eqnarray}} 
\def\nn{\nonumber} \def \bea{\begin{eqnarray}} \def\eea{\end{eqnarray}}

\newcommand{\barr}{\begin{array}}
\newcommand{\earr}{\end{array}}

  \def\G{\Gamma}
 \def\d{\delta} 

 \def\o{\omega}

\def\tb{\mc T\text{M}}

\def\mc{\mathcal}

\def\R{{\mathbb R}}

\def\one{\mbox{1 \kern-.59em {\rm l}}}

\def\bit{\begin{itemize}} 
\def\eit{\end{itemize}} 

\def\({\left(} \def\){\right)}

\begin{document}

\title{Dynamical phase space from a SO(d,d) 
matrix model}

\author{Athanasios Chatzistavrakidis}
\email{thanasis@itp.uni-hannover.de}

\affiliation{Institute for Theoretical Physics, Leibniz University Hannover, Appelstrasse 2, 30169, Hannover, Germany
}


\begin{abstract}
It is shown that a matrix model with SO($d,d$) global 
symmetry is derived from a 
generalized Yang-Mills theory on the standard Courant algebroid. This model keeps all the 
positive features of the well-studied type IIB matrix model, 
and it has many additional welcome properties. 
We show that it does not only capture the 
dynamics of spacetime, but it 
should be associated with the dynamics 
of phase space. This is supported by a large set of classical solutions of its 
equations of motion, which corresponds to phase spaces of noncommutative curved manifolds
and points to a new mechanism of emergent gravity. The model possesses a symmetry that 
exchanges positions and momenta, in analogy to quantum mechanics.
It is argued that the emergence of phase space in the model is an essential feature
for the investigation of the precise relation of matrix models 
to string theory and quantum gravity.   
\end{abstract}

\preprint{ITP-UH-11/14}

\maketitle

The concept of spacetime at very short distance scales is very different than in classical physics. Ultimately, classical spacetime 
and the gravitational field of general relativity are expected to be emergent concepts. The most prominent physical framework where this 
is indeed the case is perturbative string theory, where the starting point is an extended degree of freedom described by a non-linear 
sigma model. The perturbative quantization of the theory indeed reveals the presence of gravity. In a rather independent way, 
matrix theories \cite{Banks:1996vh,ikkt,Berenstein:2002jq} should also have something to say about quantum gravity, although the situation 
in this line of research remains more unclear. The emergence of gravity in matrix models is an interesting 
problem to address (see Ref. \cite{Steinacker:2010rh} and its references for a review of some approaches), especially since the models of Refs. \cite{Banks:1996vh,ikkt,Berenstein:2002jq} are conjectured to be 
directly related to string theory and to capture its nonperturbative dynamics.

On the other hand, it is reasonable to think that understanding the structure of spacetime at high energies is just part of the story. 
When quantum-mechanical effects become important, it can be argued that it is the structure of full phase space and its 
dynamics that would provide a more complete understanding of quantum gravity. This was emphasized recently from the point of 
view of string theory in Refs. \cite{Freidel:2013zga,Freidel:2014qna} and earlier from the point of view of 
noncommutative geometry in Refs. \cite{Madorebook,Buric:2006di,Buric:2011dd}. Given the close relation of string theory and 
noncommutative geometry \cite{Seiberg:1999vs,Connes:1997cr} and their common grounds with matrix models, it is interesting to 
examine whether the dynamics of phase space can be captured by a matrix model. In this letter we suggest such a model.
We show that starting with a generalized connection on the standard Courant algebroid we can define a Yang-Mills (YM) theory 
whose reduction to a point yields a matrix model with additional degrees of freedom and SO($d,d$) global symmetry. The symetries of 
this matrix model dictate that the classical solutions of its equations of motion (EOMs) are noncommutative phase space algebras 
that include the gravitational field, 
such as the ones described recently in Ref. \cite{Chatzistavrakidis:2014tda}. This provides an emergent picture for phase space, 
where dynamics can be incorporated and quantization can in principle be performed. 

\section*{Reductions to a point}

Let us recall that a useful way to think about matrix models is as reductions of field theories to a single point, namely to zero 
dimensions \cite{rmm1,rmm2,rmm3}. 
Consider for example the 
bosonic sector of maximal supersymmetric YM theory in 10 (Euclidean) dimensions. Its action is simply 
\be 
\int\dd^{10}x ~\sfrac 14\text{Tr}~F\w\star F~,
\ee
where 
\be 
F=\sfrac 12(\partial_MA_N-\partial_NA_M+i[A_M,A_N])\dd x^M\w\dd x^N~,
\ee
and the index $M$ takes values from 0 to 9. In order to perform a trivial dimensional reduction from 10 to 0 dimensions, we 
must assume that the gauge field in 10 dimensions does not depend on any of them, i.e. $\partial_MA_N=0$. 
Then we directly find the reduced classical bosonic action, 
\be \label{ikktaction}
S_{\text{B}}=-\sfrac 14\text{Tr}~[A_M,A_N][A_{M'},A_{N'}]g^{MM'}g^{NN'}~.
\ee
This is the starting point to define the partition function that yields the IIB matrix model \cite{ikkt},
\be \label{pf1}
\mc Z=\int \prod_{M=0}^{9} \dd A_M ~\text{Pf}(A_M)~ e^{-S_{\text{B}}}~,
\ee 
where the Pfaffian appears by integrating out the matter fields after the model is supersymmetrized.
Note that the components of the 1-form 
$A=A_M\dd x^M$ in 10D become (Hermitian) matrices in the 0D theory, having no dependence on any spacetime coordinates, which are 
anyway absent in 0 dimensions.
Of course, $A_M$ are already Hermitian matrices in 10 dimensions, since the gauge field lives in the adjoint representation 
of the gauge group. The integral in Eq. (\ref{pf1}) is over those matrices. It is remarkable that in certain cases this partition 
function, as well as similarly defined correlation functions, are convergent for the Euclidean model \cite{Austing:2001pk,Krauth:1998yu}.

The EOMs for the action (\ref{ikktaction}) are  
\be 
g^{MM'}[A_M,[A_{M'},A_N]]=0~.
\ee 
Classes of classical solutions to these equations were described in many works, such as 
the basic ones in Ref. \cite{ikkt} and more in Refs. \cite{Chatzistavrakidis:2011su,Steinacker:2011wb,Arnlind:2012cx} 
and \cite{Kim:2011cr,Kim:2011ts,Kim:2012mw} (in the Lorentzian model). 
The usual interpretation is that the matrices $A_M$ are associated to coordinates and therefore the solutions correspond to 
noncommutative spacetimes. 
This is fine, although the origin of the matrices is in the cotangent bundle and 
they naturally carry a lower index.
This remark implies that the matrices $A_M$ could also be associated to momenta and generate the momentum space 
instead of spacetime. A relevant discussion on this may be found in Ref. \cite{Aoki:1999vr}. However, there is no clear way to 
obtain the full structure of phase space from the IIB model. 
On the other hand, the momenta in 
matrix noncommutative geometry are typically related to the coordinates, since they correspond to inner derivations of the algebra 
$\mc A$ of coordinate operators \cite{Madorebook}. Moreover, they involve two copies of $\mc A$, say $\mc A_L$ and $\mc A_R$, that correspond to the left and the right action 
of the operators respectively \cite{Chatzistavrakidis:2014tda}. 
The momenta are then related to the difference $\hat x_L-\hat x_R$ of coordinate operators in the 
two representations. All these suggest that there should exist an extended model which is associated to the dynamics of phase space. 
This is desirable for the reasons explained in the introduction, primarily for a better understanding of the gravitational 
field in the framework of matrix models.

\section*{YM theories and Courant algebroids}

In order to construct the extended matrix model, we need some elementary concepts from generalized complex geometry 
\cite{Hitchin:2004ut,gg} and the theory of 
Courant algebroids \cite{wein}. The reader who is interested in the model itself may jump to the next section.
Consider the generalized tangent bundle of a manifold $\text{M}$ of dimension $d$ {\footnote{We often set $d$=10 in the 
following, although the discussion is general and holds for any $d$.}}, which is  given by 
the sum of the tangent and cotangent ones, 
$ 
\mc T\text{M}=\text{TM}\oplus\text{T}^{\star}\text{M}~.
$ 
The sections $\G(\mc T\text{M})$ of this bundle are
generalized vectors $\mathfrak{X}$, which can be written as the sum of an 1-vector 
and an 1-form, 
\be 
\mathfrak{X}=X+\eta~,\quad X\in \G(\text{TM})~,\quad \eta\in\G(\text{T}^{\star}\text{M})~.\nn
\ee
The standard Courant algebroid is obtained by equipping the above bundle with the Courant bracket \cite{dirac},
\bea 
\label{cour}[\mathfrak X,\mathfrak Y]_C=[X,Y]_{L}+{\cal L}_{X}\xi-{\cal L}_Y\eta-\sfrac 12\dd(X(\xi)-Y(\eta))~,\nn
\eea
a pairing,
\bea
\langle \mathfrak X,\mathfrak Y\rangle=\frac{1}{2}(X(\xi)+Y(\eta))~,
\eea
and a smooth map,
$ 
\rho:{\mc T}\text{M}\to\text{TM}
$,
the anchor. 
A notion with particular interest for physics is that of Dirac structures \cite{dirac}. These are vector subbundles 
$L\subset{\mc T}\text{M}$ of the generalized 
tangent bundle such that 
\bea 
\langle\mathfrak X_L,\mathfrak Y_L\rangle=0~,\quad
[\mathfrak X_L,\mathfrak Y_L]\in \G(L)~, \nn
\eea
for any $\mathfrak X_L,\mathfrak Y_L\in \G(L)$. The rank of these bundles is exactly half of the rank of $\mc T\text{M}$. 
Dirac structures are valuable for physical problems because arbitrary elements of 
$\w^{\bullet}\mc T\text{M}$ do not generically transform as tensors, however elements of $\w^{\bullet}L$ do \cite{Gualtieri:2007bq}. 
Moreover, the Courant bracket satisfies the Jacobi identity when restricted on a Dirac structure, although it does not satisfy it 
on the generalized tangent bundle.

On a vector bundle, a generalized notion of a connection can be defined \cite{Gualtieri:2007bq}. Here we consider just 
the simplest possibility,
\be \label{connection}
\mc D=\dd +A+V=\dd x^M\partial_M +A_M\dd x^M+V^M\partial_M~,
\ee
on the vector bundle $\mc T\text{M}$.
The curvature of a generalized connection is defined in a way that directly generalizes the usual definition, 
\be 
\label{gcurvgen}
 \mc F(\mathfrak X,\mathfrak Y)=[\mc D_{\mathfrak X},\mc D_{\mathfrak Y}]-\mc D_{[\mathfrak X,\mathfrak Y]}~.
\ee
 For the connection (\ref{connection}) this field 
 strength is
\bea
\mc F=&&\sfrac 12F_{MN}\dd x^M\w\dd x^N\nn\\ &&+(\partial_MV^N+i[A_M,V^N])\dd x^M\w\partial_N \nn\\ 
&&+\sfrac i2[V^M,V^N]\partial_M\w\partial_N~,
\eea
where the bracket is just the Lie algebra commutator associated to the gauge group.

Next we consider the volume form on the generalized tangent bundle. This is given as 
\bea 
\text{vol}_{\mc T\text{M}}&=&\pm\dd 
x^0\w\dots\w\dd x^9\w\partial_{0}\w\dots\w\partial_{9}~,
\eea
where the choice of sign is a choice of orientation. 
We choose the plus sign, which fixes the ordering of basis 1-forms 
and 1-vectors. 
Note that the metric does not enter, or rather the individual metric factors from 
the tangent and the cotangent bundle cancel 
each other. This becomes clear when the generalized metric
\be
{\cal H}=\begin{pmatrix}
          g-bg^{-1}b & bg^{-1} \\ 
          -g^{-1}b & g^{-1}
         \end{pmatrix}
\ee
is considered, where $g$ is a Riemannian metric on M and $b$ is a 2-form. This generalized metric transforms covariantly under O($d,d$) transformations 
$\cal O$, 
\be \label{gmtrafo}
{\cal H} \quad \to \quad {\cal O}^{T}{\cal H}{\cal O}~.
\ee
Its inverse is 
\be 
{\cal H}^{-1}=\begin{pmatrix}
          g^{-1} &  -g^{-1}b  \\ 
          bg^{-1}  & g-bg^{-1}b
         \end{pmatrix}~,
\ee
and its determinant is
$
\text{det}~{\cal H}=1,
$
thus it drops out from any relevant formula.

In order to construct a YM theory, we need a Hodge star operator on the $\mc T\text{M}$. This acts as
\bea 
\star_{\tb}:\w^p\text{TM}\w^q\text{T}^{\star}\text{M}\to \w^{d-p}\text{TM}\w^{d-q}\text{T}^{\star}\text{M}~,
\eea
and we define it such that 
$\star_{\tb}\one=\text{vol}_{\tb}$.
Applying this operation to the generalized curvature $\mc F$, we are able to compute the product $\mc F\w\star_{\tb}\mc F$ and we obtain
\bea 
\mc F\w\star_{\tb}\mc F&=&\big({\cal H}^{MM'}{\cal H}^{NN'}{\cal F}_{MN}{\cal F}_{M'N'}\big)\text{vol}_{\tb}~.\nn
\eea
 The reader should be cautious with the exhibited index structure of the 
 generalized metric, which is purely conventional
since its components 
 have both upper and lower indices. 
The expression in the parentheses can be identified with an inner product $(\mc F,\mc F)$, so that 
\be 
\mc F\w\star_{\tb}\mc F=(\mc F,\mc F) \text{vol}_{\tb}~.
\ee
The issue with this expression and the problem one faces in the corresponding generalized YM theory, is that the 
generalized curvature ${\cal F}$ does not transform as a tensor at the level of the Courant algebroid \cite{Gualtieri:2007bq}. This can be overcome by defining the theory on Dirac structures, where ${\cal F}$ transforms tensorially. This was done and examined in Ref. \cite{ChaGau}. Here we adopt a different point of view. In particular, we overcome the above problem by projecting the theory to zero dimensions, thus defining 
a matrix model, where harmful derivatives are dropped and the welcome transformation properties are restored. 

\section*{The SO(10,10) matrix model and its symmetries}

Let us first examine how the matrix model with action (\ref{ikktaction}) is obtained in this formalism. This can be approached in two 
 ways. The first way is to trace the steps that led to the type IIB matrix model. Considering the YM theory on the Dirac structure $L=\text{T}\text{M}$ of
the full Courant algebroid
and setting $b=0$, the corresponding generalized YM theory is identical 
to the standard YM in 10D and the model follows from its dimensional reduction, as previously.
Alternatively, one can consider instead the Dirac structure $L=\text{T}^{\star}\text{M}$ and the generalized YM 
theory on it. In order to reach a 0D theory, we use the technique of Refs.
\cite{Ellwood:2006ya,Mylonas:2012pg}, also used in Ref. \cite{ChaGau}, where a map to momentum space was introduced. 
Integrating out the volume of this
momentum space we obtain the action
\be \label{ikktalt}
S'_{\text{B}}=-\sfrac 14\text{Tr}~g_{MM'}g_{NN'}[V^M,V^N][V^{M'},V^{N'}]~.
\ee
This is equivalent to the action that appears in Eq. (\ref{ikktaction}) upon the identification $A_M=g_{MM'}V^{M'}$, and it 
has the same classical solutions. It is a dual model that describes the same physics. However, the two actions were obtained from two very special but different Dirac structures. Here we show that 
a more general model is obtained when we utilize the full structure of $\tb$, which has solutions that are not captured by the IIB 
matrix model. 

Consider the full generalized YM theory described in the previous section and its trivial reduction to a point. In the present case 
the 2-form $b$ is not dropped.
The result is a reduced model with bosonic action
\bea \label{action}
S&=&-\sfrac 14\text{Tr}~\biggl(\tilde{g}_{MM'}\tilde{g}_{NN'}[V^M,V^N][V^{M'},V^{N'}]\nn\\
&&+g^{MM'}g^{NN'}[A_M,A_N][A_{M'},A_{N'}]\nn\\
 &&+2~g^{MM'}\tilde{g}_{NN'}[A_M,V^N][A_{M'},V^{N'}]\nn\\
 &&-2g^{MP}g^{M'Q}b_{QN}b_{PN'}[A_M,V^N][A_{M'},V^{N'}]\nn\\
&&+2g^{MP}g^{NQ}b_{PM'}b_{QN'}[A_{M},A_{N}][V^{M'},V^{N'}]\nn\\
&&+4g^{MM'}g^{NP}b_{N'P}[A_{M},A_N][A_{M'},V^{N'}]\nn\\
&&+4g^{MP}\tilde g_{NN'}b_{M'P}[A_M,V^{N}][V^{M'},V^{N'}]\biggl)
~,
\eea
where we defined 
$ 
\tilde g=g-bg^{-1}b.
$
It should be clear that the dynamical degrees of freedom are the $A_M$ and $V^M$, while $g$ and $b$ are related to the geometry 
of the embedding space and they are not dynamical.
Note that due to the terms that appear after the first two lines,
the model is more than a simple addition of the two dual actions for the IIB model. 
Recalling the origin of the action (\ref{action}), its terms can be 
 collected accordingly. First, noting the symmetric role of $A_M$ and $V^M$, it is useful to define the extended matrix
 \be \label{collectX}
 X_M=\begin{pmatrix}
      A_M \\ V^M
     \end{pmatrix}~,
 \ee 
 where once more the position of its index is conventional and has nothing to do with its transformation properties.
 Then, the action can be cast into the following simple form:
 \be \label{collectaction}
 S=-\sfrac 14\text{Tr}~{\cal H}^{MM'}{\cal H}^{NN'}[X_M,X_N][X_{M'},X_{N'}]~.
 \ee
 A subtle point is that the bracket in 
 Eq. (\ref{collectaction}) is not precisely a commutator, since the $X_M$ are not square matrices, unlike $A_M$ and $V^M$. 
 Its actual definition is 
 \be 
 [X_M,X_N]:=\begin{pmatrix}
             [A_M,A_N] & [A_M,V^N] \\
             [V^M,A_N] & [V^M,V^N]
            \end{pmatrix}~.
 \ee

The action (\ref{action}), or equivalently (\ref{collectaction}), leads to two sets of EOMs.
Varying with respect to $A_M$ or $V^M$ independently, these are
\bea
\square A_M=0~,\quad
 \square V^M=0~,
 \eea
 where we defined the box operator 
 \bea 
 \square\cdot&=&g^{MM'}[A_M,[A_{M'},\cdot]]+\tilde{g}_{MM'}[V^M,[V^{M'},\cdot]]\nn\\&&+g^{MP}b_{M'P}\big([A_M,[V^{M'},\cdot]]
 +[V^{M'},[A_{M},\cdot]]\big)~.\nn
 \eea
 Note that these equations already appear coupled when one varies with respect to $A_M$ or $V^M$ alone.
 We are going to discuss some benchmark classical solutions in the next section.
 
 The bosonic model with action (\ref{action}) exhibits a number of symmetries. First of all, it has the obvious translational 
 symmetries $A_M\to A_M+c_M\one_{d}$ and $V^M\to V^M+c^M\one_{d}$, with $c_M, c^M\in \R$, which is an extension of the analogous property of the IIB model.
 Moreover, it has the gauge symmetry 
 $X_M\to UX_MU^{-1}$, with $U\in U(N)$, ${N}$ being the size of the matrices ($N\to\infty$, as usual for large-$N$ 
 models). This is again the same as in the IIB model and it reflects the fact that the extended set 
 of degrees of freedom originate from the same 10D generalized YM theory. Finally, there is a global rotational symmetry. 
 Recall that the Euclidean IIB model has such a symmetry too, but it is SO(10). Here we encounter the main difference, in that the model 
 (\ref{action}) exhibits a SO(10,10) global symmetry. 
 This can be directly verified by performing SO(10,10) transformations in the action (\ref{action}), 
 keeping in mind that aside $A_M$ and $V^M$, $g$ and $b$ transform too. Their transformation is 
 determined via the corresponding transformation of the generalized metric, given in Eq. (\ref{gmtrafo}).
 The model also possesses a symmetry that is not present in the IIB model, which exchanges $A_M$ and $V^M$ as
 \be
 A^M\quad \to \quad V_M\quad \text{and}\quad V_M\quad \to \quad -A^M~.\label{xp}
 \ee
 We will comment on 
 this symmetry after we present some basic classical solutions.

 \section*{Dynamical phase space}
 
 One of the prime attractive features of the IIB matrix model is that it addresses the issue of the emergence of spacetime and its dynamics 
 (see e.g. Ref. \cite{Aoki:1998vn} and Refs. \cite{Steinacker:2010rh,Nishimura:2012xs} for reviews on some recent approaches). 
 The model that we defined in the previous section is similarly the appropriate arena to study the emergence and the dynamics of 
 phase space, which is valuable for the reasons explained in the introduction. 
 
 Let us search for solutions of the classical EOMs of the model. In order to simplify our analysis, we 
 consider $b=0$ {\footnote{This has the effect of the global symmetry of the model being just SO($d$)$\times$SO($d$).}}. The general case of $b\ne 0$ is very rich and interesting and we are going to report on this is the  
 future. The EOMs simply become
 \bea
 g^{MM'}[A_M,[A_{M'},A_N]]+g_{MM'}[V^M,[V^{M'},A_N]]&=&0~,\nn\\
g_{MM'}[V^M,[V^{M'},V^N]]+g^{MM'}[A_M,[A_{M'},V^N]]&=&0~.\nn
 \eea
 Consider the following vacuum ansatz:
 \be \label{ansatz}
 A_a=\hat p_a~,\quad V^a=\hat x^a~,\quad a=1,\dots,2m,~ 2m\le d~,
 \ee
 where $\hat x^a$ and $\hat p_a$ are to be identified with position and momentum operators, and 
 $A_{2m+1}=\dots=A_d=V^{2m+1}=\dots=V^d=0$. They satisfy the 
 canonical commutation relations (CCR)
 \be
 [\hat x^a,\hat p_b]=i\hbar\d^a_b~.
 \ee
 Then the EOMs are simplified to
 \bea 
 [\hat p_a,[\hat p_{a},\hat p_b]]=0\quad \text{and} \quad
[\hat x^a,[\hat x^{a},\hat x^b]]=0~,\label{eomsimple}
 \eea
 which look very simple but actually include rather rich structures. 
 
 We split the rest of our analysis into two parts. The first part is rather degenerate, it refers to flat spacetimes and phase spaces, and 
 most of its features are essentially captured already by the IIB matrix model. It simply includes the algebra 
 \be 
 [\hat x^a,\hat x^b]=i\theta^{ab}~,\quad [\hat p_a,\hat p_b]=i\o_{ab}~,
 \ee
  with $\theta^{ab}$ and $\o_{ab}$ constant parameters, plus the CCR. This algebra is the one of noncommutative quantum mechanics 
  with a constant magnetic source \cite{Duval:2000xr,Morariu:2001dv}.
  
  The second and more interesting class of solutions contains a subset of noncommutative phases spaces recently described in Ref. \cite{Chatzistavrakidis:2014tda}. These are phase spaces of noncommutative manifolds, whose underlying
  commutative counterparts are general symplectic manifolds which are parallelizable, 
  i.e. they admit a global section of their tangent bundle, and they are not necessarily flat. It was shown in Ref. \cite{Chatzistavrakidis:2014tda} that in 
  such cases it is necessary to consider two copies of the noncommutative algebra ${\cal A}$ of position operators, one acting 
  from the left and denoted ${\cal A}_L$ with elements $\hat x_L^a$ and one acting from the right, denoted as ${\cal A}_R$ and 
  generated by $\hat x^a_R$. The two sets are commuting, namely $[\hat x^a_L,\hat x^b_R]=0$, and they are symplectic dual with respect to the symplectic 2-vector $\theta^{ab}$, i.e. $[\hat x^a_L,\hat x^b_L]=-[\hat x^a_R,\hat x^b_R]=i\theta^{ab}$. 
  In relation to the vacuum ansatz (\ref{ansatz}) fot the matrix model, $V^a$ are identified with $\hat x^a_L$, while $\hat x^a_R$ 
  do not appear explicitly in the model but only indirectly as we immediately explain.
  Recall that in the flat case,  the momentum operators act as
  \be 
 \hat p_a=\hbar\o_{ab}(\hat x^b_L-\hat x^b_R)~,
 \ee
  $\o_{ab}$ being the symplectic 2-form, and they are inner operators in the algebra $\mc A$.
  However, when the manifold is not flat these operators do not correspond to the translations generated by invariant 
  vector fields.
  In that case the correct momentum operators are
  \be 
  \hat p_i=e_i^{\ a}(\hat x_R)\hat p_a~,
  \ee 
  and this translates in the vacuum ansatz of Eq. (\ref{ansatz}) to $A_a=e_a^{\ i}\hat p_i$.
 The important aspect in this formulation is that the momenta contain the non-constant frame $e_a^{\ i}$, which is 
 associated to the 
 gravitational field. In particular, the general form of the algebra of the operators $\hat x^a$ and $\hat p_i$ turns out to be 
 \bea 
 [\hat x^a_L,\hat x^b_L]&=&-[\hat x^a_R,\hat x^b_R]=i\theta^{ab}~,\nn\\
 {[}\hat x^a_L,\hat p_i]&=&i\hbar e^a_{\ i}~,\nn\\
 {[}\hat x^a_R,\hat p_i]&=&i\hbar e^a_{\ i}-e^k_{\ b}K^{ba}_i\hat p_k~,\nn\\
{[}\hat p_i,\hat p_j]&=&M_{ij}+N_{ij}^{\ k}\hat p_k+P_{ij}^{kl}\hat p_k\hat p_l~, \label{psa}
 \eea
 with exactly computable coefficients in terms of the frame and the symplectic structure, such that 
 all the Jacobi identities are satisfied \cite{Chatzistavrakidis:2014tda}.
We observe that the gravitational field is identified with the commutation relation among the position and momentum operators, 
as in Refs. \cite{Madorebook,Buric:2006di,Buric:2011dd}. 
When the geometric data are identified with that of symplectic nilmanifolds in dimensions 4 and 6, the set of relations 
(\ref{psa}), along with the identifications $A_a=e_a^{\ i}\hat p_i$ and $V^a=\hat x^a_L$,  provides many non-trivial solutions to the equations (\ref{eomsimple}) of the model, which are not captured by the IIB matrix model. A more direct way to see this, is to 
consider the matrix model and its EOMs this time with a non-coordinate index structure. 
This happens when the starting point is a generalized coonection of the form 
\be 
{\cal D}=(\theta_I+A_I) e^I+V^I\theta_I~,
\ee
where $e^I$ and $\theta_I$ are the 1-forms and 1-vectors of the non-coordinate basis respectively. The general form of the matrix model and its EOMs remains the same in this basis, but now they are written in terms of $A_I$ and $V^I$. The Ansatz for solutions now is 
\be \label{ansatz2}
 A_i=\hat p_i~,\quad V^i=\d^i_{\ a}\hat x^a_L~,\quad i=1,\dots,2m,~ 2m\le d~.
 \ee
The EOMs in this basis become:
\bea
[\hat p_i,[\hat p_i,\hat p_j]]+[\hat x^a,[\hat x^a,\hat p_j]]&=&0~,\label{ceom1}\\
{[}\hat x^a,[\hat x^a,\hat x^b]]+[\hat p_i,[\hat p_i,\hat x^b]]&=&0~.\label{ceom2}
\eea
Assuming the phase space algebra (\ref{psa}) with constant parameters $\theta^{ab}$, we immediately obtain
\bea 
[\hat x^a,[\hat x^a,\hat p_j]]&=&[\hat x^a,i\hbar e^a_{\ j}]=0~, \nn\\  {[}\hat x^a,[\hat x^a,\hat x^b]]&=&[\hat x^a,\theta^{ab}]=0~,\nn
\eea 
where in the first equation we used the commutativity of ${\cal A}_L$ and ${\cal A}_R$. Then, a direct computation shows that the Eqs. 
(\ref{ceom1}) and (\ref{ceom2}) result in the conditions:
\begin{align}
&N_{ij}^{\ l}M_{il}+(N_{ij}^{\ l}N_{il}^{\ m}+2P_{ij}^{\ lm}M_{il})\hat p_m+\nn\\
&~~+(N_{ij}^{\ l}P_{il}^{\ mn}+2P_{ij}^{\ lm}N_{il}^{n})\hat p_m\hat p_n+\nn\\
&~~+2P_{ij}^{\ lm}P_{il}^{\ nr}\hat p_n\hat p_r\hat p_m=0~,\label{cond1}\\
&{[}\hat p_i,e^a_{\ i}]=0~.\label{cond2}
\end{align} 
Now it is time to specify a class of particular cases with their parameters. For step 2 nilmanifolds in 4 and 6 dimensions, it 
was shown in Ref. \cite{Chatzistavrakidis:2014tda} that 
\be 
M_{ij}=0,~ N_{ij}^{\ k}\propto f^k_{\ ij},~ P_{ij}^{\ kl}\propto f^k_{\ [i\underline{c}}f^l_{\ j]d}\theta^{cd}~,
\ee
while
\be 
e^a_{\ i}= \d^a_{\ i}-\sfrac 12f^a_{\ ib}\hat x_R^b~,
\ee
where $f^k_{\ ij}$ are the structure constants of the nilpotent Lie algebra that is associated to the nilmanifold. 
Then, simply using the defining relation $f^k_{\ ij}f^i_{\ lm}=0$ (no summation) for step 2 nilmanifolds, the conditions 
(\ref{cond1}) and (\ref{cond2}) are satisfied.
A full classification of solutions, including $b\ne 0$ too, is an open issue which should be addressed in detail.

We close this section by observing that the symmetry (\ref{xp}) of the matrix model translates into 
\be 
\hat x^a \quad \to \quad \hat p_a\quad \text{and} \quad \hat p_a\quad \to \quad -\hat x^a~,
\ee
which is familiar in quantum-mechanical phase space, and its role in matrix models was already emphasized in Ref. \cite{Chatzistavrakidis:2012qj}.

 \section*{Remarks on quantization}
 
Quantization in matrix models is defined via matrix integrals. For the SO(10,10) matrix model the partition function is defined as
 \be 
 {\cal Z}=\int \prod_{M=0}^{9} \dd A_M\prod_{N=0}^{9} \dd V^N ~ e^{-S}~,
 \ee
 where $S$ is given by Eq. (\ref{action}). Correlation functions may be defined similarly.
A primary question is whether these integrals are convergent 
 under certain conditions. This is a technical issue which presents an interesting challenge. However, given that when $V^N$ vanish
 the corresponding integrals  are convergent for certain number of dimensions (including 10) and 
 certain gauge groups \cite{Austing:2001pk,Krauth:1998yu}, it is reasonable to expect that a careful evaluation will 
 reveal such cases for the extended model too. This will be addressed in future work.

 \section*{Conclusions}
 
 In the present work we argued that a better understanding of the dynamics of full phase space, rather than just spacetime, 
 can be relevant for physics at the Planck scale and ultimately for quantum gravity. Similar ideas were already emphasized 
 before \cite{Freidel:2014qna,Madorebook}. Here we constructed a theory that captures the dynamics of phase space. 
 It is given by a matrix model which extends in a consistent way previous matrix models that proved to be 
 successful in the description of spacetime dynamics \cite{Banks:1996vh,ikkt}. The model is derived from the trivial dimensional 
 reduction of a generalized Yang-Mills theory on a Courant algebroid to zero dimensions. This allows us to overcome the problem of 
 the nontensorial transformation of generalized fields on the Courant algebroid. The symmetries of the model include and 
 extend the ones 
 of the IIB model. Notably there is a global SO($d,d$) symmetry, as well as a quantum-mechanical symmetry that is interpreted as 
 exchange of positions and momenta in phase space. Certain noncommutative phase spaces that correspond to curved manifolds 
 are classical solutions of the EOMs. The key feature is that the commutator of positions and momenta can be associated to the gravitational field, and therefore (semiclassical) gravity naturally emerges on solutions of the model. Furthermore, quantization is in principle 
 possible, with the partition function and correlation functions defined via matrix integrals. Whether these integrals are 
 convergent remains an open issue which should be carefully addressed.

\paragraph*{Acknowledgments.~}  The author is indebted to F. F. Gautason and L. Jonke for reading the manuscript and making many interesting 
comments, as well as to H. Steinacker for useful remarks. This work was completed in the Simons Center for Geometry and Physics during the 2014 Simons Summer Workshop. The author thanks the organisers for the 
excellent working environment and the center for financial support.

\end{document}